\documentclass{article}
\usepackage{graphicx}
\usepackage{wrapfig}
\usepackage{subfigure}
\usepackage{booktabs}
\usepackage{color}
\usepackage{cancel}
\usepackage{soul}

\title{\textbf{\large{When Viscous Jets Collide;\\Liquid Chains, Threads, Webs, Fishbones and Balloons.}}}
\author{\textsc{\small{Bavand Keshavarz$^*$, and Gareth H. McKinley}}\\ \small{\textit{Department of Mechanical Engineering, Massachusetts Institute of Technology}}\\ \small{\textit{77 Massachusetts Avenue, Cambridge - MA 02139, USA}}\\ {\small $^*$Corresponding Author: bavand@mit.edu}}
\date{October, 14, 2013}

\begin{document}
\maketitle

\begin{abstract} 
In this fluid dynamics video prepared for the APS-DFD Gallery of Fluid Motion we study the collision of two identical viscous jets at different jet speeds and collision eccentricities. The dynamics of the jet motion are slowed down by orders of magnitude using a synchronized strobe effect coupled with precise timing control of the perturbation frequency imposed on one of the jets. Our results show that different shapes and morphologies appear as we change the collision eccentricity. At low jet speeds ($We_j = 45$) viscous threads and filaments are formed as the jets begin to impinge on each other.  As the propagation axis of one of the jets (jet A) is moved closer to the center of the other stream (jet B) they exert a torque on each other.  Because of the resulting swirl, jet A rotates around the axis of jet B and finally breaks into droplets through the action of capillary forces; the pinch off of droplets is similar to a stream of balloons which are released in the air.  At zero eccentricity the two jets unite into one jet at low speed and the unified jet breaks up after undergoing Rayleigh-Plateau instability.  For higher jet velocities ($We_j=125$) different morphologies develop; at high eccentricities of impact (0.95), threads will appear; triangular sheets or webs form at eccentricities close to 0.9 and as one jet moves closer to the center of the other one first partial/incomplete fishbones appear and eventually periodically-dancing fishbones develop.  
\end{abstract}

\end{document}